\documentclass[twocolumn,showpacs,preprintnumbers,amsmath,amssymb,aps]{revtex4}
\usepackage{graphicx}
\usepackage{dcolumn}
\usepackage{bm}
\begin{document}
\preprint{}
\title{ Effect of isospin asymmetry in nuclear system}
\author{Shailesh K. Singh} %\email{shailesh@iopb.res.in}
\author{S. K. Biswal} %\email{sbiswal@iopb.res.in}
\author{M. Bhuyan} %\email{bunuphy@iopb.res.in}
\author{S. K. Patra} %\email{patra@iopb.res.in}
\affiliation{Institute of Physics, Sachivalaya Marg, Bhubaneswar -
751005, India.} 
\begin{abstract}
The effect of $\delta-$ and $\omega-\rho-$meson cross couplings on asymmetry 
nuclear systems are analyzed in the frame-work of an effective Field theory 
motivated 
relativistic mean field formalism. The calculations are done on top of the 
G2 parameter set, where these contributions are absent. We calculate the 
root mean square radius, binding energy, single particle energy 
(for the $1^{st}$ and
last occupied orbits), density and spin-orbit interaction potential for 
some selected nuclei and 
evaluate the $L_{sym}-$ and $E_{sym}-$ coefficients for nuclear matter as 
function
of $\delta-$ and  $\omega-\rho-$meson coupling strengths. As expected, 
the influence of these effects are negligible for symmetry nuclear system and
these effects are very important for systems with large isospin asymmetry.

\end{abstract}
\maketitle

\section{Introduction}
In recent years the effective field theory approach to quantum hadrodynamic 
(QHD) has been studied extensively. The parameter set G2 
\cite{Fu96,Fu97}, 
obtained from the effective field theory motivated Lagrangian (E-RMF) 
approach, is very successful in reproducing the nuclear matter properties 
including the structure of neutron star as well as of finite nuclei 
\cite{arumugam}. This model well reproduce the experimental values of 
binding energy, root mean square (rms) radii and other finite nuclear 
properties \cite{patra01,patra02,patra03}. Similarly, the prediction of 
nuclear matter properties including the phase transition as well as the 
properties of compact star are remarkably good \cite{patra04,patra05}. 
%Identical to the linear relativistic mean field (RMF) model of Walecka
%\cite{walecka01} and the nonlinear RMF model \cite{boguta01}, the
%E-RMF approximation gives the spin-orbit interaction in an $ab initio$
%manner. 
The G2 force parameter is the largest force set available, 
in the relativistic mean field model. It contains almost all interaction 
terms of nucleon with mesons, self and cross coupling of mesons upto 
$4^{th}$ order.

In the effective field theory motivated
relativistic mean field (E-RMF) model of Furnstahl et al
\cite{Fu96,Fu97}, the coupling of $\delta-$meson is not taken into account.
Also, the effect of $\rho$ and $\omega$ meson cross coupling was
neglected. It is soon realized that the importance of $\delta$ 
meson \cite{kubis} and the cross coupling of $\omega$ and $\rho-$mesons 
\cite{gmuca} can not
be neglected while studying the nuclear and neutron matter
properties. 
Horowitz and Piekarewicz \cite{horo01} studied explicitly the
importance of $\rho$ and  $\omega$ cross coupling
to finite nuclei as well as to the properties of neutron star
structures.  This coupling also influences the
nuclear matter properties, like symmetry energy $E_{sym}$, slope parameters 
$L_{sym}$ and curvature $K_{sym}$ of $E_{sym}$ \cite{shailesh13}. It is shown in
Ref. \cite{arumugam} that the self- and cross couplings of $\omega$ meson
plays an important role to make the nuclear equation of state
(EOS) softer.

The observation of Brown \cite{brown00} and later on by Horowitz and 
Piekarewicz \cite{horo01} make it clear that the neutron radius of heavy
nuclei have a direct correlation with the equation of state (EOS) of 
compact star matter. 
It is shown that the collection of neutron to proton radius difference 
$\triangle r=r_n-r_p$ using relativistic and nonrelativistic formalisms show
two different patterns. Unfortunately, the error bar in neutron radius 
makes no difference between these two pattern. Therefore, the experimental
result of JLAB \cite{jlab} is much awaited. 
To have a better argument for all this, 
Horowitz and Piekarewicz \cite{horo01} introduced $\Lambda_s$ and 
$\Lambda_v$ couplings to take care of the skin thickness in
$^{208}$Pb as well as the crust of neutron star. The symmetry energy, 
and hence the neutron radius, plays an important
role in the construction of asymmetric nuclear EOS. Although, the new couplings
 $\Lambda_s$ and $\Lambda_v$ take care of the neutron radius problem, the
effective mass splitting between neutron and proton is not taken care. This 
effect can not be neglected in a highly neutron-rich dense matter system
and drip-line nuclei. In addition to this mass splitting, the rms charge
radius anomaly of $^{40}$Ca and $^{48}$Ca may be resolved by this
scalar-isovector $\delta-$meson inclusion to the E-RMF model. 
Our aim in this paper is to see the effect of $\delta-$ and $\rho-\omega-$mesons
couplings in a highly asymmetric system, like asymmetry finite nuclei, 
neutron star and asymmetric EOS.  

The paper is organized as follows: First of all we extended the 
E-RMF Lagrangian by including the $\delta-$meson and the
$\omega-\rho$ cross couplings. The field equations are derived
from the extended Lagrangian for finite nuclei. Then the
equation of state for nuclear matter and neutron star
matters are derived. 
The calculated results are discussed in section III. In this section,
we study the effect of $\delta-$meson on asymmetric nuclear matter, including
the neutron star. Then, we adopt the calculations for finite nuclei and
see the changes in binding energy, radius etc. In the
last section, the conclusions are drawn.

\section{Formalism }

The relativistic treatment of the quantum hadrodynamic (QHD) models 
automatically include the
spin-orbit force, the finite range and the density dependence of the
nuclear interaction. The non-linearity of the $\sigma-$meson coupling 
included the 3-body interaction \cite{shief}, which is currently noticed 
as an important 
ingredient for nuclear saturation. The relativistic mean field (RMF) or
the E-RMF model has the advantage that, with the
proper relativistic kinematics and with the meson properties already
known or fixed from the properties of a small number of finite nuclei,
it gives excellent results for binding energies, root-mean-square
radii, quadrupole and hexadecapole deformations and other properties
of  spherical and deformed nuclei \cite{Re86,Sh93,Ga90,Ru88,Pa91}. The
quality of the results is comparable to that found in non-relativistic
nuclear structure calculations with effective Skyrme \cite{Va72} or
Gogny \cite{De80} forces.

The theory and the equations for finite nuclei
and nuclear matter can be found in Refs. \cite{Se97,Fu96,Fu97,Se96} 
and we shall only outline the formalism
here. We start from Ref. \cite{Fu96} where the field equations were
derived from an energy density functional containing Dirac baryons and
classical scalar and vector mesons. Although this energy functional
can be obtained from the effective Lagrangian in the Hartree
approximation \cite{Se97,Fu97}, it can also be considered as an
expansion in terms of ratios of the meson fields and their gradients
to the nucleon mass. The energy density functional for finite nuclei 
can be written as \cite{Se97,Fu97,Se96}:
\begin{eqnarray}
{\cal E}({r}) & = &  \sum_\alpha \varphi_\alpha^\dagger({r})
\Bigg\{ -i \mbox{\boldmath$\alpha$} \!\cdot\! \mbox{\boldmath$\nabla$}
+ \beta [M - \Phi({r}) - \tau_3 D(r)] \nonumber \\
&+& W({r})
+ \frac{1}{2}\tau_3 R({r})
+ \frac{1+\tau_3}{2} A ({r}) \nonumber \\
%\nonumber \\[3mm]
%& &
&-& \frac{i \beta\mbox{\boldmath$\alpha$}}{2M}\!\cdot\!
  \left (f_v \mbox{\boldmath$\nabla$} W({r})
  + \frac{1}{2}f_\rho\tau_3 \mbox{\boldmath$\nabla$} R({r}) \right)
  \Bigg\} \varphi_\alpha (r)
\nonumber \\[3mm]
%& & \null
&+& \left ( \frac{1}{2}
  + \frac{\kappa_3}{3!}\frac{\Phi({r})}{M}
  + \frac{\kappa_4}{4!}\frac{\Phi^2({r})}{M^2}\right )
   \frac{m_s^2}{g_s^2} \Phi^2({r})
%-  \frac{\zeta_0}{4!} \frac{1}{ g_v^2 } W^4 ({r}) 
\nonumber \\
%\nonumber \\[3mm]
%& & \null 
&+& \frac{1}{2g_s^2}\left( 1 +
\alpha_1\frac{\Phi({r})}{M}\right) \left(
\mbox{\boldmath $\nabla$}\Phi({r})\right)^2  
-  \frac{\zeta_0}{4!} \frac{1}{ g_v^2 } W^4 ({r}) \nonumber \\
&-& \frac{1}{2g_v^2}\left( 1 +\alpha_2\frac{\Phi({r})}{M}\right)
% \nonumber \\[3mm]
% & &  \null 
 \left( \mbox{\boldmath $\nabla$} W({r})  \right)^2
 - \frac{1}{2e^2} \left( \mbox{\boldmath $\nabla$} A({r})\right)^2
 \nonumber \\[3mm]
% & &  \null 
&-& \frac{1}{2}\left(1 + \eta_1 \frac{\Phi({r})}{M} +
 \frac{\eta_2}{2} \frac{\Phi^2 ({r})}{M^2} \right)
  \frac{m_v^2}{g_v^2} W^2 ({r})
   \nonumber \\[3mm]
 &-& \null
     \frac{1}{2g_\rho^2} \left( \mbox{\boldmath $\nabla$} R({r})\right)^2
   - \frac{1}{2} \left( 1 + \eta_\rho \frac{\Phi({r})}{M} \right)
   \frac{m_\rho^2}{g_\rho^2} R^2({r})
\nonumber\\[3mm]
   &+&
    \frac{1}{2 g_{\delta}^{2}}\left( \mbox{\boldmath $\nabla$} D({r})\right)^2
   -\frac{1}{2}\frac{ {m_{\delta}}^2}{g_{\delta}^{2}}\left(D^{2}(r)\right)
\nonumber \\
&-&
   \Lambda_{v}\left(R^{2}(r)\times W^{2}(r)\right),
\label{eqFN1}
\end{eqnarray}
% Equation 1
%
where $\Phi$, $W$, $R$, $D$ and $A$ are the fields for 
$\sigma, \omega, \rho, \delta$ and photon and $g_\sigma$, $g_\omega$,
$g_\rho$, $g_\delta$ and $\frac{e^2}{4\pi}$ are their coupling constant, 
respectively. The masses of the mesons are $m_\sigma$, $m_\omega$, $m_\rho$
and $m_\delta$ for $\Phi_0$, $V_0$, $b_0$ and $\delta_0$, respectively.
In the energy functional, the non-linearity as well as the cross-coupling
upto a maximum of $4^{th}$ order is taken into account. This is restricted 
due the condition $1 > \frac{field}{M}$ (M = nucleon mass) and 
non-significant contribution of the higher order \cite{patra01}. The higher 
non-linear coupling for $\rho-$ and $\delta-$meson fields are not taken in 
the  energy functional, because the expectation values of the $\rho-$ 
and $\delta-$ fields are order of magnitude less than that of 
$\omega-$field and they have only marginal contribution to finite nuclei.
%
%where the index $\alpha$ runs over all occupied states of the positive
%energy spectrum, $\Phi \equiv g_s \phi_0$, $ W \equiv g_v V_0$, $R
%\equiv g_\rho b_0$ and $A \equiv e A_0$ and $D \equiv g_{\delta}\delta$. 
%Except for the terms with
%$\alpha_1$ and $\alpha_2$, the functional (\ref{eqFN1}) is of fourth
%order in the expansion. Following Refs.\ \cite{Se97,Fu97,Es99}, we
%retain the fifth-order terms $\alpha_1$ and $\alpha_2$ because their
%contribution to the nuclear surface energy is numerically of the same
%magnitude as the contribution from the quartic scalar term. One can
%see that the new terms concentrate on the isoscalar channel and that
%the expansion with respect to the isovector meson is shorter (the
%$\eta_\rho$ coupling is of third order). Higher non-linear couplings
%of the $\rho$ meson are not considered because the expectation value
%of the $\rho$ field is typically an order of magnitude smaller than
%that of the $\omega$ field \cite{Se97,Fu97}, and they only have a
%marginal impact on the usual properties studied for terrestrial
%nuclei. 
For example, in calculations of the high-density equation of
state, M\"uller and Serot \cite{Se96} found the effects of a quartic
$\rho$ meson coupling ($R^4$) to be appreciable only in stars made of
pure neutron matter. A surface contribution $-\alpha_3 \Phi \, (
\mbox{\boldmath$\nabla$} R )^2 /(2 g_\rho^2 M)$ was tested in Ref.\
\cite{Es99} and it was found to have absolutely negligible effects.
We should note, nevertheless, that very recently it has been shown
that couplings of the type $\Phi^2 R^2$ and $W^2 R^2$ are useful to
modify the neutron radius in heavy nuclei while making very small
changes to the proton radius and the binding energy \cite{horo01}.

The Dirac equation corresponding to the energy density eqn. (\ref{eqFN1})
becomes
\begin{eqnarray}
\Big\{&-&i \mbox{\boldmath$\alpha$} \!\cdot\! \mbox{\boldmath$\nabla$}
+ \beta [M - \Phi(r) - \tau_3 D(r)] + W(r) % \nonumber \\
+ \frac{1}{2} \tau_3 R(r) \nonumber \\
&+& \frac{1 +\tau_3}{2}A(r)
%\nonumber \\[3mm]
% \null  
\left.
   -  \frac{i\beta \mbox{\boldmath$\alpha$}}{2M}
   \!\cdot\! \left [ f_v \mbox{\boldmath$\nabla$} W(r)
    + \frac{1}{2}f_{\rho} \tau_3 \mbox{\boldmath$\nabla$} R(r) \right]
     \right\} \varphi_\alpha (r) \nonumber \\
&=& 
     \varepsilon_\alpha \, \varphi_\alpha (r).%\nonumber 
     \label{eqFN12}
     \end{eqnarray}

The mean field equations for $\Phi$, $W$, $R$, $D$ and $A$ are given by
\begin{eqnarray}
-\Delta \Phi(r) + m_s^2 \Phi(r)  &=&
g_s^2 \rho_s(r)
-{m_s^2\over M}\Phi^2 (r) 
\left({\kappa_3\over 2}+{\kappa_4\over 3!}{\Phi(r)\over M}
\right )
\nonumber  \\[3mm]
&+& \null
{g_s^2 \over 2M}
\left(\eta_1+\eta_2{\Phi(r)\over M}\right)
{ m_v^2\over  g_v^2} W^2 (r)
\nonumber  \\[3mm]
&+& \null
 {\eta_{\rho} \over 2M}{g_s^2 \over g{_\rho}^2}
{ m_\rho^2 } R^2 (r)
+ {\alpha_2 \over 2M} {g_s^2\over g_v^2}
(\mbox{\boldmath $\nabla$}W(r))^2 \;\; 
\nonumber  \\[3mm] 
&+& \null
 {\alpha_1 \over 2M}[
(\mbox{\boldmath $\nabla$}\Phi(r))^2
+2\Phi(r)\Delta \Phi(r) ]
 \label{eqFN2}  \\[3mm]
-\Delta W(r) +  m_v^2 W(r)  & = &
g_v^2 \left( \rho(r) + \frac{f_v}{2} \rho_{\rm T}(r) \right)
-{1\over 3!}\zeta_0 W^3(r)
\nonumber  \\[3mm]
&-& \null
 \left( \eta_1+{\eta_2\over 2}{\Phi(r)\over M} \right ){\Phi(r)
\over M} m_v^2 W(r)
\nonumber  \\[3mm]
&+& \null
 {\alpha_2 \over M} [\mbox{\boldmath $\nabla$}\Phi(r)
\cdot\mbox{\boldmath $\nabla$}W(r)
%                    +\Phi(r)\Delta W(r)] \,,
+\Phi(r)\Delta W(r)]
\nonumber\\
&-&
 {2}\Lambda_{v}{g_{v}}^2 {R^{2}(r)} W(r) \,,
\label{eqFN3}  \\[3mm] 
-\Delta R(r) +  m_{\rho}^2 R(r)  & = &
{1 \over 2 }g_{\rho}^2 \left (\rho_{3}(r) +
{1 \over 2 }f_{\rho}\rho_{\rm T,3}(r) \right ) 
\nonumber\\
&-&
   \eta_\rho {\Phi (r) \over M }m_{\rho}^2 R(r)
-{2}\Lambda_{v}{g_{\rho}}^{2} R(r) {W^{2}(r)} \,,
\label{eqFN4}  \\[3mm] 
-\Delta A(r)   & = &
e^2 \rho_{\rm p}(r),    \\ 
-\Delta D(r)+{m_{\delta}}^{2} D(r) & = &
g_{\delta}^{2}\rho_{s3}, 
\label{eqFN5a}
\end{eqnarray}
% Equation 2,3,4,5
%
where the baryon, scalar, isovector, proton and tensor densities are
%
% Equations 6,7,8,9,10,11
\begin{eqnarray}
\rho(r) & = &
\sum_\alpha \varphi_\alpha^\dagger(r) \varphi_\alpha(r) \,,
\label{eqFN6} \\[3mm]
\rho_s(r) & = &
\sum_\alpha \varphi_\alpha^\dagger(r) \beta \varphi_\alpha(r) \,,
\label{eqFN7} \\[3mm]
\rho_3 (r) & = &
\sum_\alpha \varphi_\alpha^\dagger(r) \tau_3 \varphi_\alpha(r) \,,
\label{eqFN8} \\[3mm]
\rho_{\rm p}(r) & = &
\sum_\alpha \varphi_\alpha^\dagger(r) \left (\frac{1 +\tau_3}{2}
\right)  \varphi_\alpha(r) \,,
\label{eqFN9}  \\[3mm]
\rho_{\rm T}(r) & = &
\sum_\alpha \frac{i}{M} \mbox{\boldmath$\nabla$} \!\cdot\!
\left[ \varphi_\alpha^\dagger(r) \beta \mbox{\boldmath$\alpha$}
\varphi_\alpha(r) \right] \,,
\label{eqFN10} \\[3mm]
\rho_{\rm T,3}(r) & = &
\sum_\alpha \frac{i}{M} \mbox{\boldmath$\nabla$} \!\cdot\!
\left[ \varphi_\alpha^\dagger(r) \beta \mbox{\boldmath$\alpha$}
\tau_3      \varphi_\alpha(r) \right], \,  
\label{eqFN11}  \\[3mm]
\rho_{s3} (r) & = & \sum_{\alpha}\varphi_{\alpha}^\dagger (r)\tau_{3}\beta
\varphi_{\alpha} (r), 
\end{eqnarray}
%$\qquad$     
where $\rho_{s3}$=$\rho_{sp}-\rho_{sn}$, $\rho_{sp}$ and $\rho_{sn}$ 
are scalar densities for proton and neutron respectively.  
The scalar density $\rho_{s}$ is expressed as the sum of 
proton(p) and neutron(n) densities $\rho_{s}$=$\langle\psi 
\psi\rangle$=$\rho_{sp}$+$\rho_{sn}$, which are given by
\begin{equation}
\rho_{s i}=\frac{2}{(2\pi)^3}\int_{0}^{k_{i}} d^{3}k \frac{M_{i}^{\ast}}
{(k^{2}+M_{i}^{\ast 2})^{\frac{1}{2}}}, \qquad i= p,n
\end{equation}
$k_{i}$ is the nucleon's Fermi momentum and $M_{p}^{\ast}$, $M_{n}^{\ast}$
are the proton and neutron effective masses, respectively and can be written as
\begin{equation}
M_{p}^{\ast}=M-g_{s}\phi _{0}-g_{\delta}\delta,
\end{equation}
\begin{equation} 
M_{n}^{\ast}=M-g_{s}\phi _{0}+g_{\delta}\delta.
\end{equation}
Thus, the $\delta$ field splits the nucleon effective masses.
The baryon density is given by
\begin{equation}
\rho_{B}=\langle\psi\gamma^{0}\psi\rangle=\gamma\int_{0}^{k_{F}}\frac{d^{3}k}
{(2\pi)^{3}},
\end{equation}
where $\gamma$ is spin or isospin multiplicity ($\gamma=4 $ for symmetric 
nuclear matter and $\gamma = 2$ for pure neutron matter). 
The proton and neutron Fermi momentum will also split, while they have to 
fulfill the following condition:
\begin{eqnarray}
\rho_{B}&=&\rho_{p}+\rho_{n}
\nonumber\\
& &
=\frac{2}{(2\pi)^{3}}\int_{0}^{k_{p}}d^{3}k
+\frac{2}{(2\pi)^{3}}\int_{0}^{k_{n}}d^{3}k.
\end{eqnarray}
Because of the uniformity of the nuclear system for infinite nuclear 
matter all of the gradients of the fields in
Eqs.\ (\ref{eqFN1})--(\ref{eqFN5a}) vanishes  and only the $\kappa_3$,
$\kappa_4$, $\eta_1$, $\eta_2$ and $\zeta_0$ non-linear couplings
remain. Due to the fact that the solution of symmetric 
nuclear matter
in mean field depends on the ratios $g_s^2/m_s^2$ and $g_v^2/m_v^2$
\cite{Se86}, we have seven unknown parameters. By imposing the values
of the saturation density, total energy, incompressibility modulus and
effective mass, we still have three free parameters (the value of
$g_\rho^2/m_\rho^2$ is fixed from the bulk symmetry energy coefficient
$J$).
The energy density and pressure of nuclear matter is given by
\begin{eqnarray}\label{eqn:eos1}
\epsilon & = &  \frac{2}{(2\pi)^{3}}\int d^{3}k E_{i}^\ast (k)+\rho (r) W(r)
+\frac{1}{2}\rho_{3}(r)R(r)
 \nonumber \\
& & 
+\frac{ m_{s}^2\Phi^{2}}{g_{s}^2}\Bigg(\frac{1}{2}+\frac{\kappa_{3}}{3!}
\frac{\Phi (r)}{M} + \frac{\kappa_4}{4!}\frac{\Phi^2({r})}{M^2}\Bigg)
-\frac{1}{4!}\frac{\zeta_{0}W^{4}(r)}{g_{v}^2}
\nonumber\\
&& 
 -\frac{1}{2}m_{v}^2\frac{W^{2}(r)}{g_{v}^2}\Bigg(1+\eta_{1}\frac{\Phi}{M}+\frac{\eta_{2}}{2}\frac{\Phi ^2}{M^2}\Bigg) 
+\frac{1}{2}\frac{m_{\delta}^2}{g_{\delta}^{2}}\left(D^{2}(r) \right)
 \nonumber\\
 &&
-\frac{1}{2}\Bigg(1+\frac{\eta_{\rho}\Phi(r)}{M}\Bigg)\frac{m_{\rho}^2}{g_{\rho}^2}R^{2}(r)
-\Lambda_{v}R^{2}(r)\times W^{2}(r),
\end{eqnarray}
\begin{eqnarray}\label{eqn:eos2}
P & = &  \frac{2}{3 (2\pi)^{3}}\int d^{3}k \frac{k^2}{E_{i}^\ast (k)}-
\frac{ m_{s}^2\Phi^{2}}{g_{s}^2}\Bigg(\frac{1}{2}+\frac{\kappa_{3}}{3!}
\frac{\Phi (r)}{M}+ \frac{\kappa_4}{4!}\frac{\Phi^2({r})}{M^2}  \Bigg)
\nonumber\\ 
& &
 +\frac{1}{2}m_{v}^2\frac{W^{2}(r)}{g_{v}^2}\Bigg(1+\eta_{1}\frac{\Phi}{M}+\frac{\eta_{2}}{2}\frac{\Phi ^2}{M^2}\Bigg)+\frac{1}{4!}\frac{\zeta_{0}W^{4}(r)}{g_{v}^2}
  \nonumber\\
& &
+\frac{1}{2}\Bigg(1+\frac{\eta_{\rho}\Phi(r)}{M}\Bigg)\frac{m_{\rho}^2}{g_{\rho}^2}R^{2}(r)+\Lambda_{v}R^{2}(r)\times W^{2}(r)
  \nonumber\\
& &
-\frac{1}{2}\frac{m_{\delta}^2}{g_{\delta}^{2}}\left(D^{2}(r)\right),
\end{eqnarray}
where $E_{i}^\ast (k)$=$\sqrt {k^2+{M_{i}^\ast}^2} \qquad  (i= p,n)$.
In the context of density functional theory, it is possible to
parametrize the exchange and correlation effects through local
potentials (Kohn--Sham potentials), as long as those contributions be
small enough \cite{Ko65}. The Hartree values are the ones that control the
dynamics in the relativistic Dirac-Br\"corner-Hartree-Fock (DBHF)
 calculations. Therefore, the local meson
fields in the RMF formalism can be interpreted as Kohn--Sham potentials 
and in this sense equations (\ref{eqFN2})--(\ref{eqFN5a}) include effects 
beyond the Hartree approach through the non-linear couplings 
\cite{Se97,Fu96,Fu97}.

\newpage
\section{RESULTS AND DISCUSSIONS}

Our calculated results are shown in Figs. ($1-10$)  and Table I for both 
finite nuclei and infinite nuclear matter systems. The effect of 
$\delta-$meson and the crossed coupling constant $\Lambda_v$ of $\omega-\rho$ 
fields on some selected nuclei like $^{48}$Ca and $^{208}$Pb are demonstrated 
in Figs. $1-4$ and the nuclear matter outcomes are displayed in rest of the 
figures and table. In one of our recent  publication \cite{shailesh13}, 
the  explicit dependence of  $\Lambda_v(\omega-\rho)$ on nuclear matter 
properties are shown and it is found that it has significant implication 
on various physical properties, like mass and radius of neutron star and
$E_{sym}$ asymmetry energy and its slope parameter $L_{sym}$ for 
infinite nuclear matter system at high densities. Here, only the
influence of $\Lambda_v$ on finite nuclei and that of $g_{\delta}$ 
on both finite and infinite nuclear systems are studied.  

\subsection{Finite Nuclei}
\begin{figure}
\caption{\label{fig1}Binding energy (BE), root mean square radius 
and first ($1s^{n,p}$) and last ($1f^n$, $2s^p$) occupied orbits for 
$^{48}$Ca as a function of $g_{\delta}$ and $\Lambda_v$.}
\includegraphics[width=1.0\columnwidth]{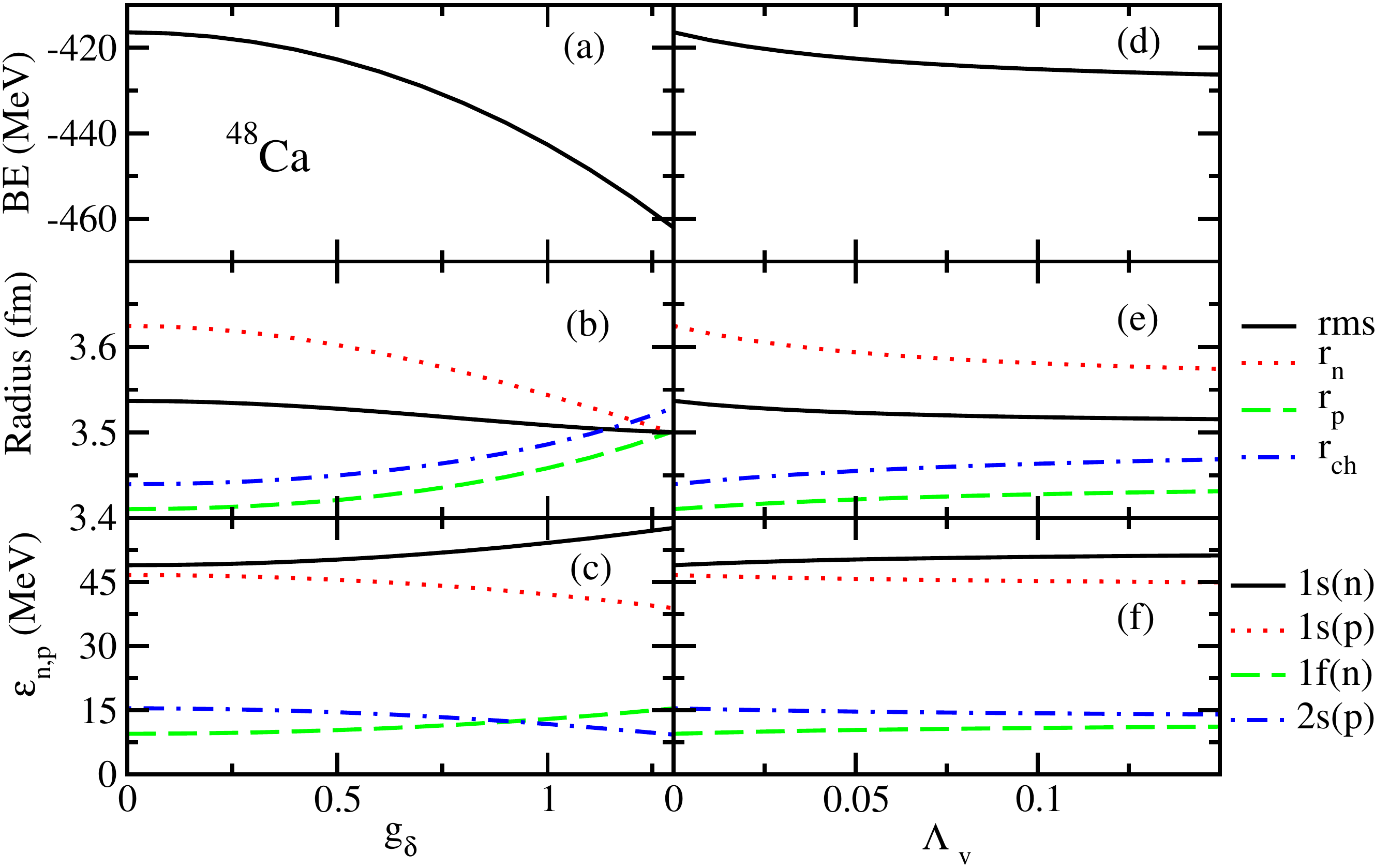}
\end{figure}

\begin{figure}
\caption{\label{fig2}Same as Fig. \ref{fig1} for
$^{208}$Pb.}
\includegraphics[width=1.0\columnwidth]{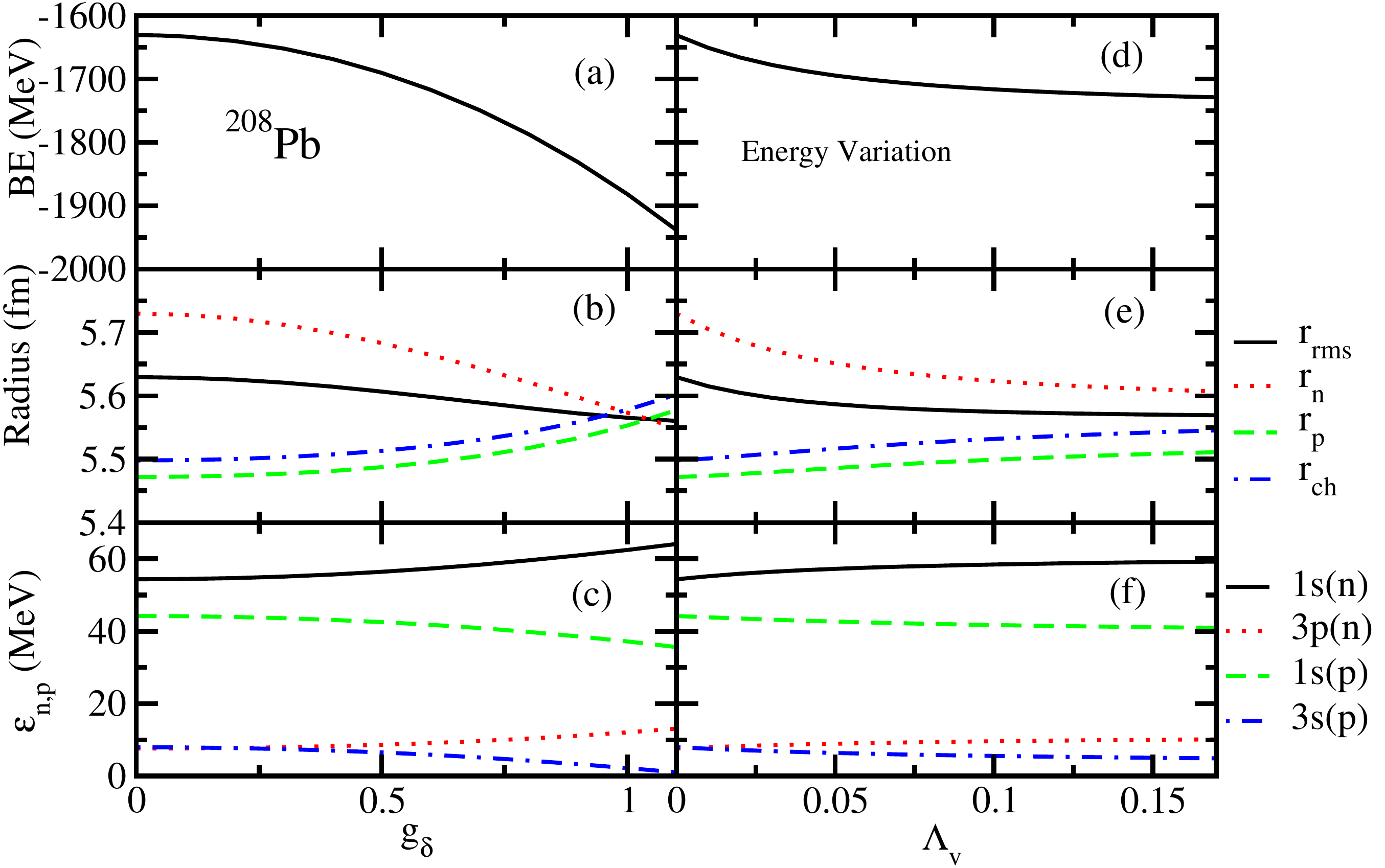}
\end{figure}

In this section we analyzed the effects of $\delta$ meson 
and $\Lambda_v$ coupling in finite nuclei. For this, we calculate the 
binding energy (BE), rms radii ($r_n$, $r_p$, $r_{ch}$, $r_{rms}$), 
and energy of first and last filled orbitals 
of $^{48}$Ca and $^{208}$Pb with $g_{\delta}$ and $\Lambda_v$. 
The finite size of the nucleon 
is taken into account for the charge radius using the relation
$r_{ch}=\sqrt{r_p^2+0.64}$. The results are shown in Figs. 
\ref{fig1}, \ref{fig2}.

In our calculations, while 
analyzing the effect of $g_{\delta}$, we keep $\Lambda_v=0$ and 
vice versa. From the figures, it is evident that the binding energy, radii and
single particle levels $\epsilon_{n,p}$ affected drastically with $g_{\delta}$
contrary to the effect of $\Lambda_v$. A careful inspection shows a slight
decrease of $r_n$ with the increase of $\Lambda_v$ consistent with the
analysis of \cite{mario10}.  Again, it is found that the binding energy
increases with increasing of the coupling strength
upto $g_\delta \sim 1.5$ and no convergence solution 
available beyond this value. Similar to the $g_{\delta}$ limit, there is 
limit for $\Lambda_v$ also, beyond which no solution exist.
From the anatomy of $g_\delta$ on $r_n$ and $r_p$, we find their
opposite trend in size. That means the value of $r_n$ decreases and
$r_p$ increases with $g_{\delta}$ for both $^{48}$Ca and $^{208}$Pb. It
so happens that both the radii meet at a point near $g_\delta=1.0$
(Fig 1 and Fig. 2) and again shows reverse character on
increasing $g_\delta$, i.e., the neutron skin thickness ($r_n-r_p$) 
changes its sign with $g_\delta$.
This interesting results may help us to settle the charge radius anomaly 
of $^{40}$Ca and $^{48}$Ca. 

In Fig. 1(c), we have shown the first ($1s^{n,p}$) and last 
($1f^n$ and $2s^p$) filled orbitals for $^{48}$Ca as a function 
of $g_{\delta}$ and $\Lambda_v$. The effect of $\Lambda_v$ is marginal,
i.e., almost negligible on $\epsilon_{n,p}$ orbitals. However, this is significance
with the increasing value of $g_{\delta}$. The top most filled orbital
even crosses each other at $g_{\delta} \sim 1$, although initially, it is
well separated. On the other hand, the first filled orbital $1s$ both
for proton and neutron get separated more and more with $g_{\delta}$, which
has almost same single particle energy $\epsilon_{n,p}$ at $g_{\delta}=0$.
We get similar trend for $^{208}$Pb, which is shown in Fig. 2(c). In both the
representative cases, we notice orbital flipping only for the last filled
levels.

\begin{figure}
\caption{\label{fig3} The neutron, proton and total density with radial 
coordinate $r(fm)$ at different values of $g_\delta$ (a) and 
$\Lambda_v$ (c). The variation of spin-orbit potential for proton and neutron 
are shown in (b) and (d) by keeping the same $g_\delta$ and $\Lambda_v$
as (a) and (c) respectively.   }
\includegraphics[width=1.0\columnwidth]{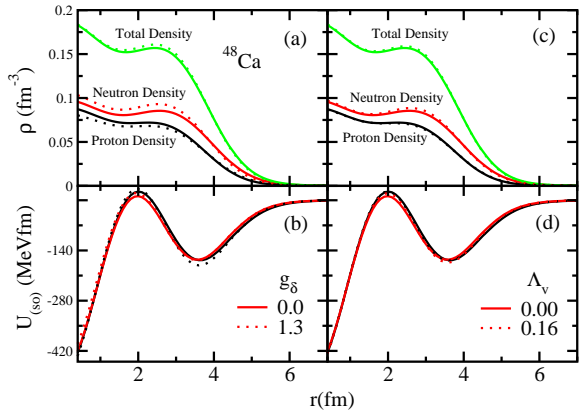}
\end{figure}

\begin{figure}
\caption{\label{fig4} Same as Fig. \ref{fig3} for $^{208}$Pb.}
\includegraphics[width=1.0\columnwidth]{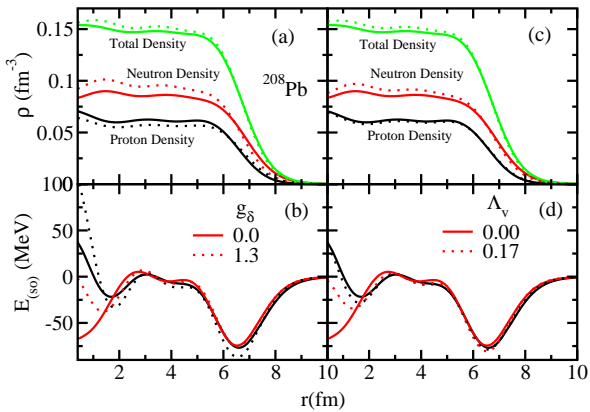}
\end{figure}

The nucleon density distribution (proton $\rho_p$ and neutron $\rho_n$) 
and spin orbit interaction potential $U_{so}$ of finite nuclei 
are shown in Figs. \ref{fig3} and \ref{fig4}. The calculations are done 
with two different values of $g_\delta$ and $\Lambda_v$ as shown in 
the figures. Here, the solid line is drawn for initial  
and dotted one is for the limiting values.  
In Fig. \ref{fig3}(a), we have depicted the neutron, proton and total 
density distribution for $^{48}$Ca at values of $g_\delta=0.0$ and 1.3.
Comparing Figs. \ref{fig3}(a) and \ref{fig3}(c), one can see 
that the sensitivity of $g_\delta$ is more than $\Lambda_v$ on density 
distribution.  The spin-orbit potential $U_{so}$ of $^{48}$Ca with different 
values of $g_\delta$ are shown in Fig. \ref{fig3}(b) and for $\Lambda_v$ 
in Fig. \ref{fig3}(d).  Similarly, we have given these observables for 
$^{208}$Pb in Fig. \ref{fig4}.  
In general, for light mass region both coupling constants $g_{\delta}$ and 
$\Lambda_v$ are less effective in density distribution and spin-orbit 
potential.  It is clear from this analysis that the coupling strength of 
$\delta-$meson is more influential than the isoscalar-vector and 
isovector-vector cross coupling.
This effect is mostly confined to the central region of the nucleus. 

\subsection{Nuclear Matter}

In this section, we do calculation for
nuclear matter properties like energy and pressure densities, symmetry energy,
radii and mass of the neutron star using $\omega-\rho$ and $\delta$ couplings
on top of G2 parametrization.
Recently, it is reported \cite{shailesh13} that the $\omega-\rho$ cross 
coupling plays a vital role for nuclear matter system on important 
physical observables like equation of state, symmetry energy coefficient,
$L_{sym}$ coefficient etc. A detail account is available in Ref. 
\cite{shailesh13} for $\omega-\rho$ coupling on nuclear matter system.
The main aim of this section is to take $\delta$
meson as an additional degree of freedom in our calculations and elaborate 
the effect on nuclear matter system within G2 parameter set. 
In highly asymmetric
system like neutron star and supernova explosion, the contribution of 
$\delta$ meson is important. This is because of the high asymmetry due
to the isospin as well as the difference in neutron and
proton masses.
Here, in the calculations the $\beta-$equilibrium and charge 
neutrality conditions are not considered. We only varies the neutron and
proton components with an asymmetry parameter $\alpha$, defined as
$\alpha=\frac{\rho_n-\rho_p}{\rho_n+\rho_p}$. The splitting in nucleon masses 
is evident from equations (16) and (17) due to the inclusion of isovector 
scalar $\delta-$meson. For $\alpha$=0.0, the nuclear matter system is
purely symmetrical and for other non-zero value of $\alpha$, the
system get more and more asymmetry. For $\alpha=1.0$, it is a case of pure
neutron matter.

In Fig. \ref{fig5}(a), the effective masses of proton and neutron 
are given as a function of $g_\delta$. As we have mentioned, $\delta-$meson is
responsible for the splitting of effective masses (Eqns. (16) and (17)),
this splitting increases continuously 
with coupling strength $g_\delta$. In Fig. \ref{fig5}, the splitting is
shown for few representative cases at $\alpha$=0.0, 0.75 and 1.0.
The solid line is for  $\alpha$=0.0 and $\alpha$=0.75, 1.0 are 
shown by dotted and dashed line, respectively. From the figure, it is
clear that the effective mass is unaffected for symmetric matter.
The proton effective mass $M_p^*$ is above the reference line with 
$\alpha=0$ and the neutron effective mass always lies below it.
The effect of $g_\delta$ on binding energy 
per nucleon is shown in Fig. \ref{fig5}(b) and pressure density 
in Fig. \ref{fig5}(c). One can easily see the effect of $\delta$ 
meson interaction on the energy and pressure density of the nuclear system.
The energy and pressure density show opposite trend to each other
with the increase function of $g_{\delta}$.

\begin{figure}
\caption{\label{fig5} Variation of nucleonic effective masses, binding energy
per particle (BE/A) and pressure density as a function of $g_{\delta}$
on top of G2 parameter set for nuclear
matter.}
\includegraphics[width=1.0\columnwidth]{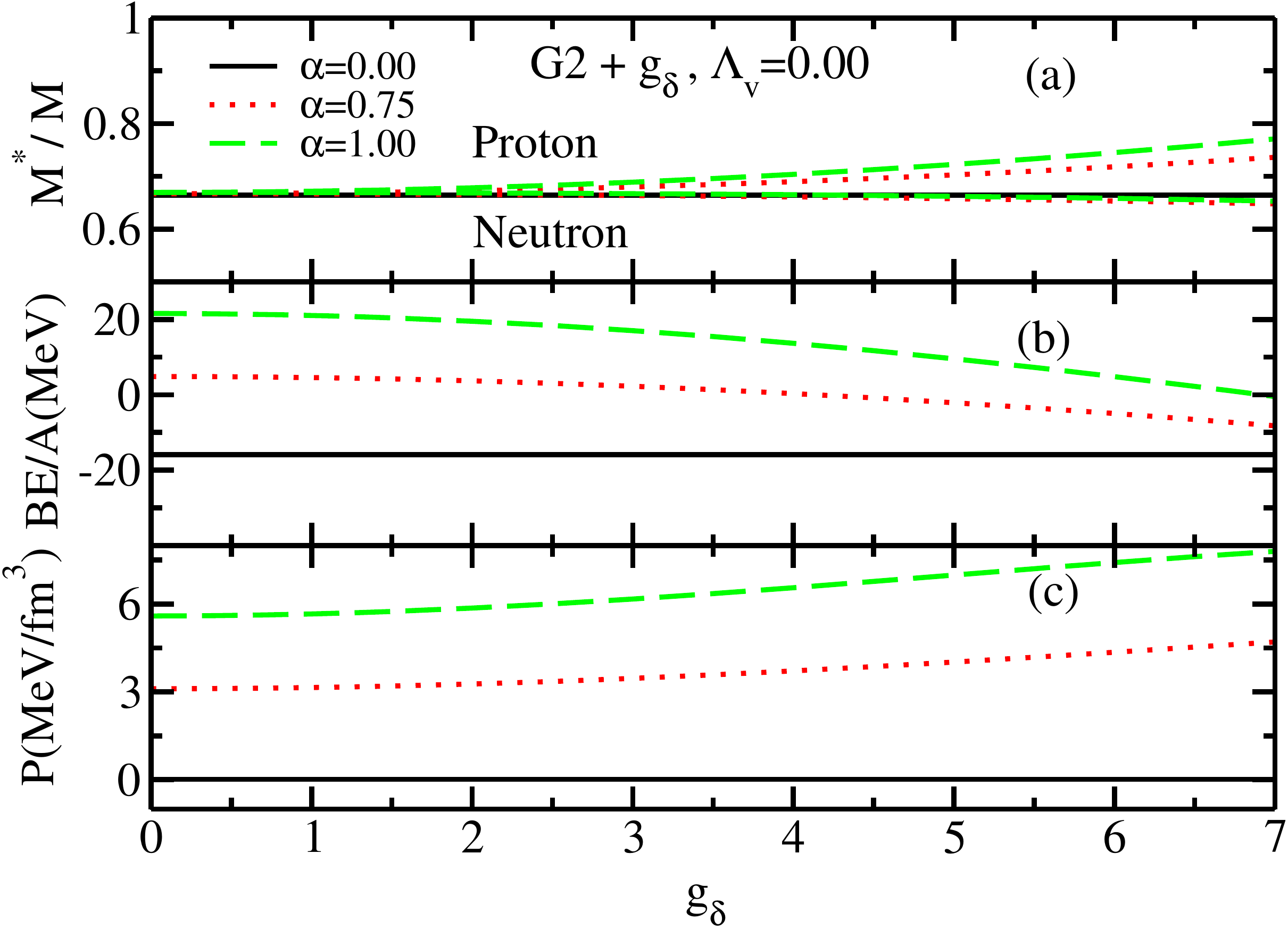}
\label{fig5}
\end{figure}

\subsection{Energy and Pressure Density}

\begin{figure}
\caption{\label{fig6} Energy per particle and pressure density with 
respect to density with various $g_\delta$.}
\includegraphics[width=1.0\columnwidth]{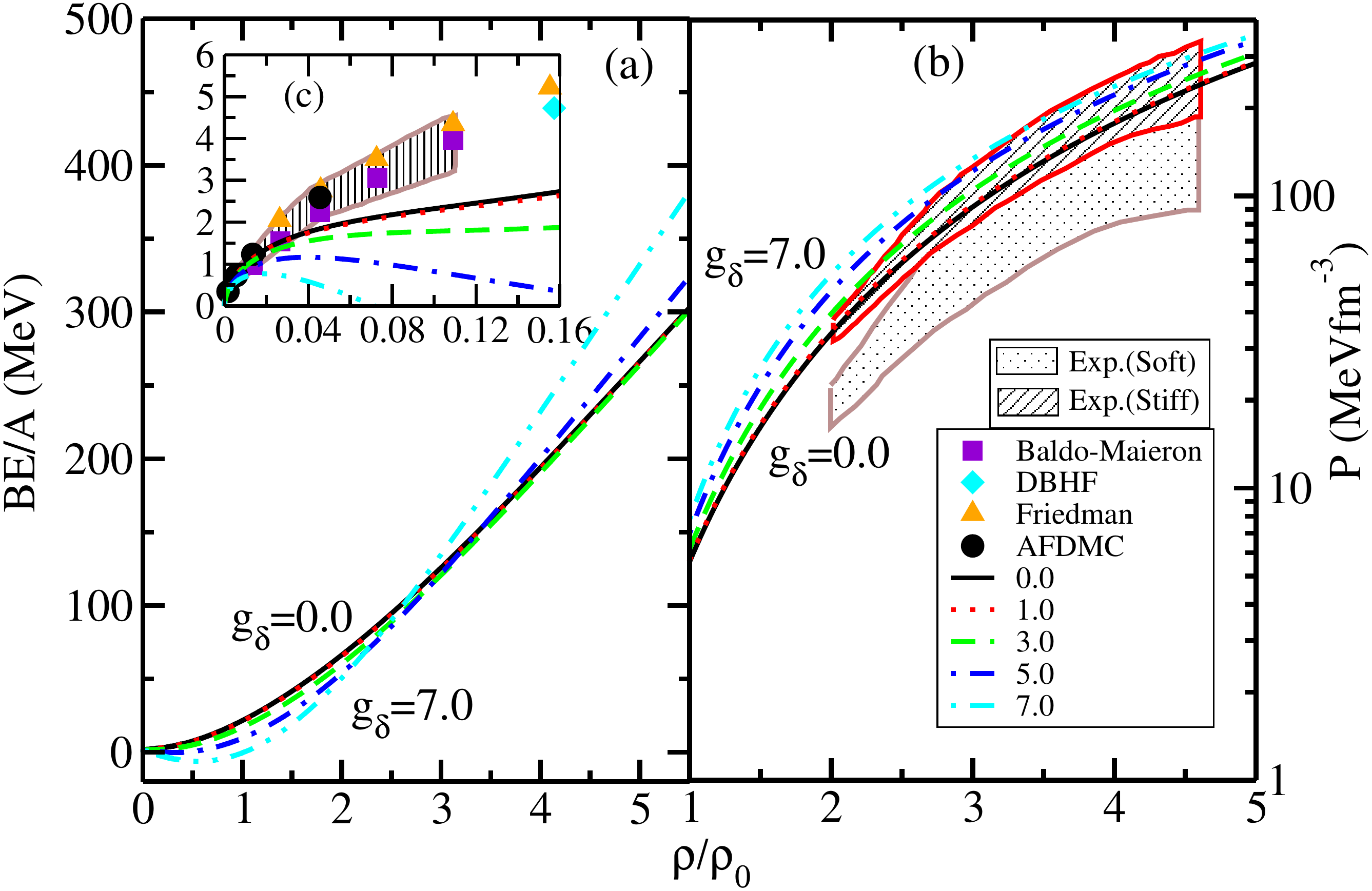}
\end{figure}

We analyze the binding energy per nucleon and pressure 
density including the contribution of $\delta-$meson in the G2 
Lagrangian as a function of density. As it is mentioned earlier, the 
addition of $\delta-$meson is done due to its importance on asymmetry
nuclear matter as well as to make a full fledge E-RMF model.
This is tested by calculating the observables at  
different values of $\delta-$meson coupling strength $g_\delta$. 
In Fig. \ref{fig6}, the calculated BE/A and ${\cal P}$  for 
pure neutron matter with baryonic density for different  
$g_\delta$ are shown. Unlike to the small value of $g_{\delta}$ upto
1.5 in finite nuclei, the instability arises at $g_\delta$=7.0 in
nuclear matter. Of course, this limiting value of $g_\delta$
depends on the asymmetry of the system.

In Fig. \ref{fig6}(a), we have given BE/A for different 
values of $g_\delta$. 
It is seen from Fig. \ref{fig6}(a), the binding increases with $g_\delta$ 
in the lower density region and maximum value of 
binding energy is $\sim$ 7 MeV for $g_{\delta}$=7.0. On the
other hand, in higher density region, the binding energy curve 
for finite $g_{\delta}$ crosses the one with $g_{\delta}$=0.0.
That means, the EOS with $\delta-$meson is stiffer than the one
with pure G2 parametrization. As a result, one get a heavier mass
of the neutrons star, which suited with the present experimental finding
\cite{demorest2010}.
For comparing the data at lower density 
(dilute system, 0 $< \rho/\rho_0 < $ 0.16) the zoomed version of the
region is shown as an inset Fig.
\ref{fig6}(c) inside Fig. \ref{fig6}(a). From the
zoomed inset portion, it is clearly seen that the curves 
with various $g_{\delta}$ at $\alpha=1.0$ (pure neutron matter)
deviate from other theoretical 
predictions, such as Baldo-Maieron \cite{baldo08}, 
DBHF \cite{margu07}, Friedman \cite{fried81}, auxiliary-field 
diffusion Monte
Carlo (AFDMC) \cite{gando08} and Skyrme interaction \cite{shf12}.    
This is an inherited problem in the RMF or E-RMF formalisms, which need
more theoretical attention. Similarly, the  
pressure density for different values of 
$g_\delta$ with G2 
parameter set are given in Fig. \ref{fig6}(b). At high 
density we can easily see that the curve becomes more stiffer with 
 the coupling strength $g_{\delta}$.
The experimental constraint of equation
of state obtained from heavy ion flow data for both stiff and soft 
EOS is also displayed for comparison in the region 2 $< \rho/\rho_0 <$ 4.6
\cite{daniel02}. Our results match with the stiff EOS data of 
Ref. \cite{daniel02}.

\subsection{Symmetry Energy}

The symmetric energy $E_{sym}$ is important in infinite
nuclear matter and finite nuclei, because of isospin dependence in
the interaction. The isospin asymmetry
arises due to the difference in densities and masses of the neutron and proton,
respectively. The density type isospin asymmetry is taken care by $\rho $ meson
(isovector-vector meson) and mass asymmetry by $\delta-$meson 
(isovector - scalar meson).
The expression of symmetry energy $E_{sym}$ is a combine expression of 
$\rho-$ and $\delta-$mesons, which is defined
as \cite{matsui81,kubis,patra01,roca11}:

\begin{eqnarray}
E_{sym}(\rho)=E_{sym}^{kin}(\rho) + E_{sym}^{\rho}({\rho})+E_{sym}^{\delta}
(\rho),
\end{eqnarray}

with

\begin{eqnarray}
E_{sym}^{kin}(\rho)=\frac{k_F^2}{6E_F^*};\;
E_{sym}^{\rho}({\rho})=\frac{g_{\rho}^2\rho}{8m_{\rho}^{*2}}
\end{eqnarray}

and 

\begin{eqnarray}
E_{sym}^{\delta}(\rho)=-\frac{1}{2}\rho \frac{g_\delta^2}{m_\delta^2}\left(
\frac{m^*}{E_F}\right)^2 u_\delta \left(\rho,m^*\right).
\end{eqnarray}
The last function $u_\delta$ is from the discreteness of the 
Fermi momentum. This momentum is quite large in nuclear matter system and
can be treated as a continuum and continuous system.
The function $u_\delta$ is defined as:

\begin{eqnarray} 
u_\delta \left(\rho,m^*\right)=\frac{1}{ 1+ 3 \frac{g_\delta^2}{m_\delta^2}}
\left(\frac{\rho^s}{m^*}-\frac{\rho}{E_F}\right).
\end{eqnarray}
In the limit of continuum, the function $u_\delta \approx 1$. 
The whole symmetry energy ($E_{sym}^{kin}+E_{sym}^{pot}$) 
arises from $\rho-$ and $\delta-$mesons  is given as:
\begin{eqnarray}{\label{eqn:sym}}
E_{sym}(\rho)=\frac{k_F^2}{6E_F^*}+\frac{g_{\rho}^2\rho}{8{m_{\rho}^{*2}}}  
-\frac{1}{2}\rho \frac{g_\delta^2}{m_\delta^2}\left(\frac{m^*}{E_F}\right)^2 
u_\delta \left(\rho,m^*\right),
\end{eqnarray}
where the effective energy $E_F^*={\sqrt{(k_F^2+m^{*2})}}$, $k_F$ is the
Fermi momentum and the effective mass $m^*=m-g_s\phi_0\pm g_{\delta}\delta_0$. 
The effective mass of the $\rho$-meson modified, because of cross coupling
of $\rho-\omega$ and is given by
\begin{eqnarray}
m_{\rho}^{*2}=\left(1+\eta_{\rho}\frac{g_{\sigma}\sigma}{m_B}\right)m_{\rho}^2
+2g_{\rho}^2(\Lambda_vg_v^2\omega_0^2).
\end{eqnarray}
The cross coupling of isoscalar-isovector mesons ($\Lambda_v$) 
modified the density dependent of $E_{sym}$ without affecting the saturation
properties of the symmetric nuclear matter (SNM). This is explained
explicitly in Ref. \cite{shailesh13} and no need special attention here.
In E-RMF model with pure G2 set, the symmetric nuclear matter saturates 
at $\rho_0$ = 0.153$fm^{-3}$, $BE/A$ = 16.07 MeV, compressibility 
$K_0$ = 215 MeV and symmetry energy of $E_{sym}$= 36.42 MeV 
\cite{Fu96,Fu97}.

\begin{figure}
\caption{\label{fig7}Symmetry energy $E_{sym}$ (MeV) of neutron 
matter with respect to different value of $g_\delta$ on top of G2 parameter 
set. The heavy ion collision (HIC) experimental data 
\cite{HIC} (shaded region) and non-relativistic Skyrme GSkII  \cite{gskii}, 
and Skxs20 \cite{skxs20} predictions are also given.
$\Lambda_v$=0.0 is taken.}
\includegraphics[width=1.0\columnwidth]{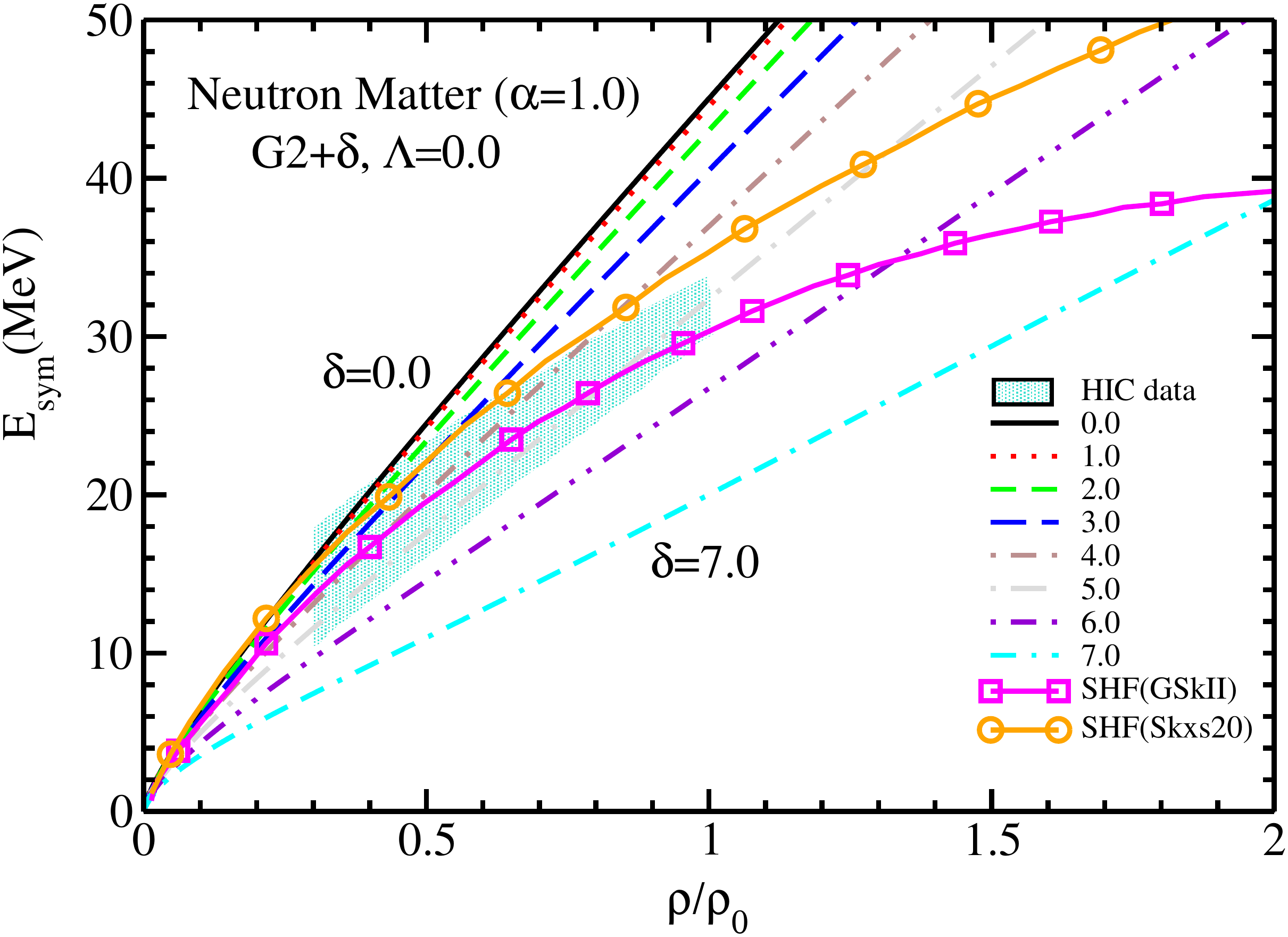}
\end{figure}

In the numerical calculation, the coefficient of symmetry energy $E_{sym}$ 
is obtained by the energy difference of symmetry and pure neutron matter 
at saturation and it is defined by Eqn. (\ref{eqn:sym}) for a quantitative 
description at various densities.  
Our results for  $E_{sym}$ are compared in Fig. \ref{fig7} with  
experimental heavy ion collision (HIC) data \cite{HIC} and other theoretical 
predictions of non-relativistic Skyrme-Hartree-Fock model.  
The calculation is done for pure neutron matter with different values of 
$g_{\delta}$, which are compared with two selective force parameter sets 
GSkII \cite{gskii} and Skxs20 \cite{skxs20}.
For more discussion one can see Ref. \cite{shf12}, where 240 different 
Skyrme parametrizations are used. Here in our calculation, as usual 
$\Lambda_v=0$  to see the effect of $\delta-$meson coupling on $E_{sym}$.    
In this figure, shaded region represent the HIC data \cite{HIC} within 
$0.3 < \rho/\rho_0 < 1.0$ region 
and the symbols square and circle represent the SHF results for GSkII and 
Skxs20 respectively. 
Analysing Fig. \ref{fig7}, $E_{sym}$ 
of G2 matches with the shaded region in low density region, however as 
the density increases, the value of $E_{sym}$ moves away. 
Again, the symmetry energy becomes softer by 
increasing the value of coupling strength $g_\delta$. For higher value of 
$g_\delta$, again the curve moves far from the empirical shaded area. 
In this way, we can 
fix the limiting constraint on coupling strength of $\delta-$ meson and nucleon.
Similar to the finite nuclear case, the nuclear matter system becomes
unstable for excessive value of $g_\delta$ ($>7.0$). 
This constrained may help to improve the G2+$g_{\delta}$ parameter set for both
finite and infinite nuclear systems.

\begin{figure}
\caption{\label{fig8a}Symmetry energy $E_{sym}$ (MeV), slope coefficients $L_{sym}$ (MeV) 
and $K_{sym}$ (MeV) at different $g_{\delta}$ with $\Lambda_v$=0.0.}
\includegraphics[width=1.0\columnwidth]{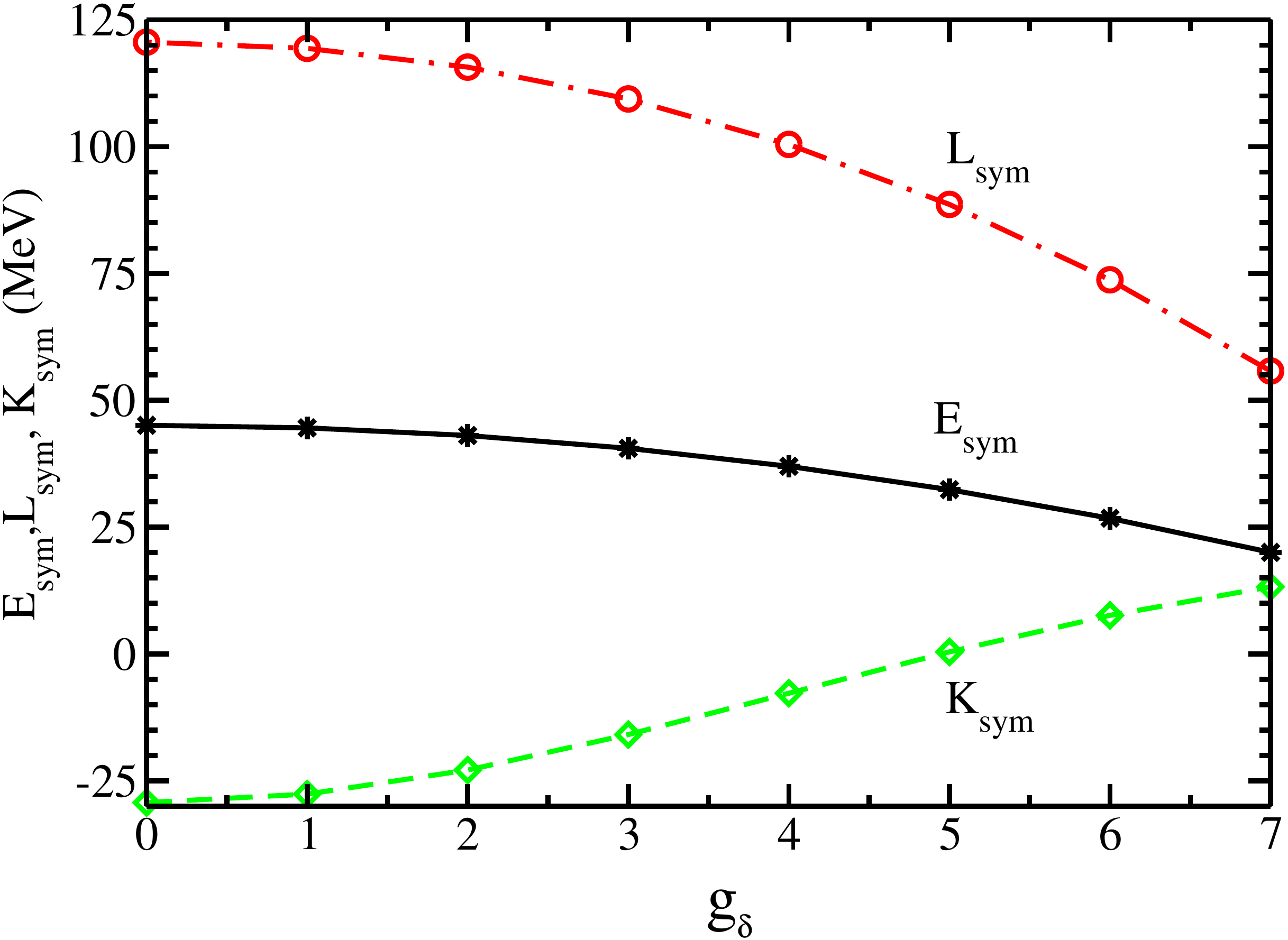}
\end{figure}
\begin{figure}
\caption{\label{fig8} Constraints on $E_{sym}$ with its first derivative, 
i.e., $L_{sym}$ at saturation density for neutron matter. The 
experimental results of HIC \cite{HIC}, PDR \cite{pdr07,pdr10} and IAS 
\cite{ias}
are given. The theoretical prediction of finite range droplet model (FRDM)
and Skyrme parametrization are also given \cite{frdm12}, SHF \cite{shf12}. }
\includegraphics[width=1.0\columnwidth]{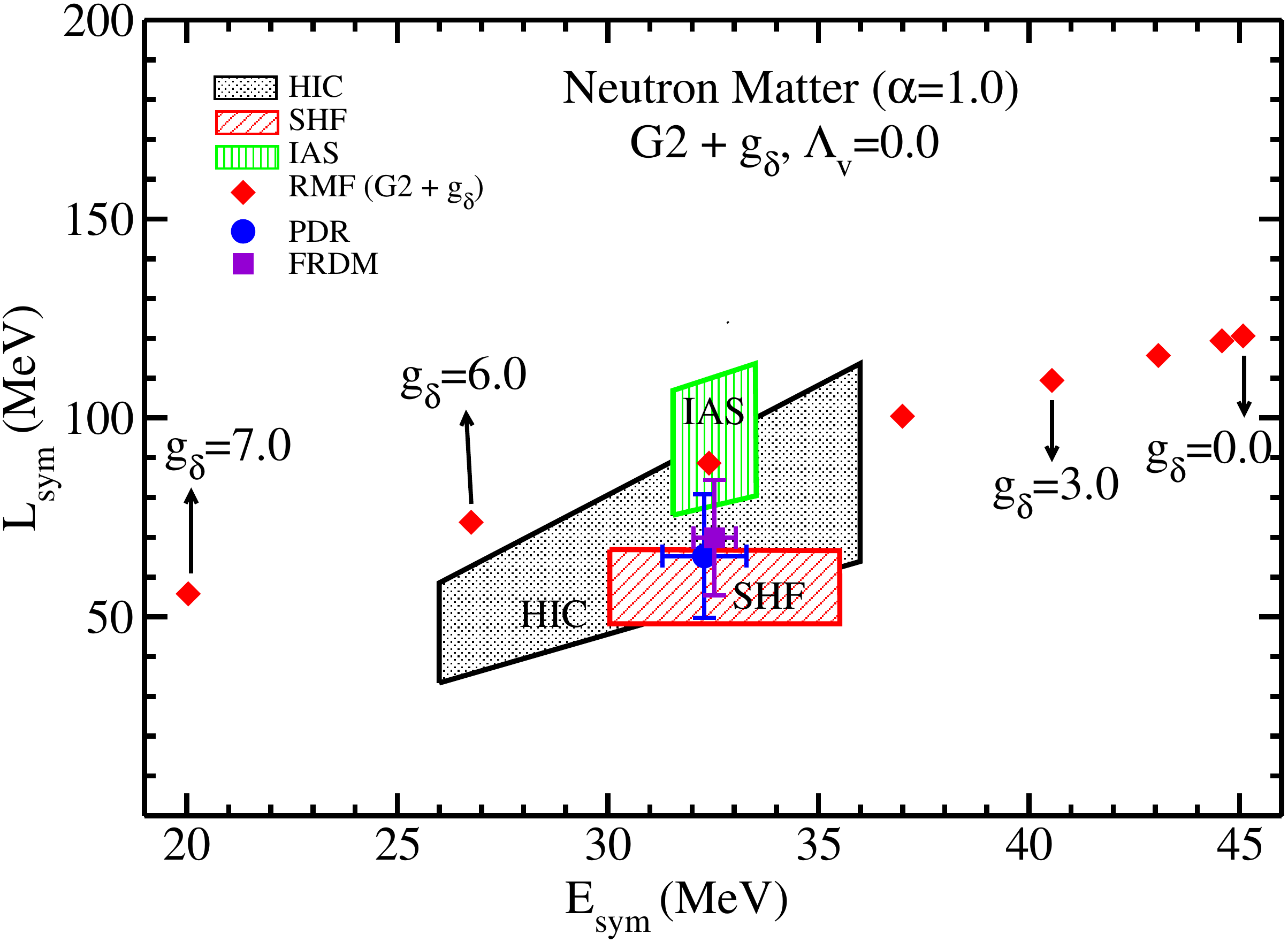}
\end{figure}

The symmetry energy of a nuclear system is a function of baryonic density $\rho$,
hence can be expanded in a Taylor series around the 
saturation density $\rho_0$ as (\ref{eqn:sym}):
\begin{eqnarray}
E_{sym}(\rho)=E_0 + L_{sym} {\cal Y} + \frac{1}{2}K_{sym} {\cal Y}^2 + O[{\cal Y}^3],
\end{eqnarray} 
where $E_0=E_{sym}(\rho=\rho_0)$, ${\cal Y} = \frac{\rho-\rho_0}{3\rho_0}$ and
the coefficients $L_{sym}$ and $K_{sym}$ are defined as:
\begin{eqnarray}
L_{sym}=3\rho\left(\frac{\partial E_{sym}}{\partial \rho}
\right)_{\rho=\rho_0},\;
K_{sym}=9\rho^2\left(\frac{\partial^2 E_{sym}}{\partial\rho^2}\right)
_{\rho=\rho_0}.
\end{eqnarray}
%r
Here $L_{sym}$ is the slope parameter defined as the slope of 
$E_{sym}$ at saturation. The quantity $K_{sym}$ represents the
curvature of $E_{sym}$ with respect to density.
A large number of investigation have been made to fix the value of
$E_{sym}$, $L_{sym}$ and $K_{sym}$ \cite{shailesh13,HIC,shf12,xu10,new11,ste12,fatt12}.
In Fig. \ref{fig8}, we have given the symmetry energy with its first 
derivative at saturation density with different values of 
coupling strength staring from $g_{\delta}= 0.0-7.0$. The variation
of $E_{sym}$, $L_{sym}$ and $K_{sym}$ with $g_{\delta}$ 
are listed in Table \ref{tab:tab1}. 
The variation in symmetry energy takes place from 45.09 to 20.04 MeV, 
$L_{sym}$ from 120.60 to 55.78 MeV and $K_{sym}$ from $-29.28$ to 13.27 MeV
at saturation density corresponding to $0.0 < g_\delta < 7.0$.
From this investigation, one can see that G2 set is not sufficient to predict this 
constrained on $E_{sym}$ and $L_{sym}$. It is suggestive to introduce the $\delta-$ 
meson as an extra degree of freedom into the model to bring the data within the
prediction of experimental and  other theoretical constraints.  

\begin{table}
\renewcommand{\tabcolsep}{0.5cm}
\renewcommand{\arraystretch}{1.0}
\caption{\label{tab:tab1}The symmetry energy 
$E_{sym}$ (MeV), slope co-efficient $L_{sym}$ (MeV) and $K_{sym}$ (MeV) at 
different values of $g_\delta$.}
\begin{tabular}{ccccc}
\hline
$g_\delta$ &  $E_{sym}$ & $L_{sym}$ & $K_{sym}$  \\
\hline
0.0 & 45.09 & 120.60 & -29.28 \\
1.0 & 44.58 & 119.37 & -27.61 \\
2.0 & 43.07 & 115.67 & -22.87 \\
3.0 & 40.55 & 109.41 & -15.87 \\
4.0 & 37.00 & 100.44 & -7.72  \\
5.0 & 32.40 & 88.61  & 0.43   \\
6.0 & 26.76 & 73.77  & 7.61   \\
7.0 & 20.04 & 55.78  & 13.27  \\
\hline
\end{tabular}
\end{table}
The above tabulated results are also depicted in Fig. \ref{fig8a}
to get a graphical representation of $E_{sym}$, $L_{sym}$ and $K_{sym}$.
The values of  $E_{sym}$ is marginally effective with the $\delta-$meson
coupling strength. However, in the same time $L_{sym}$ and $K_{sym}$
vary substantially as shown in the figure. The slope parameter $L_{sym}$
decreases almost exponentially opposite to the similar exponential increase
of $K_{sym}$. At large value of $g_{\delta}$ all the three quantities
almost emerge very closely to the similar region.

\subsection{Neutron Star}

In this section, we study the effect of $\delta-$meson
on mass and radius of neutron star. Recently, experimental observation predicts the 
constraint on mass of neutron star and its radius \cite{demorest2010}. 
This observation suggests that
the theoretical models should predict the star mass  
and radius as $M\geq(1.97\pm0.04)M_\odot$ and $11 < R(km) < 15$.
Keeping this point in mind, we calculate the mass
and radius of neutron star and analyzed their variation with $g_{\delta}$.

In the interior part of neutron star, the neutron chemical potential exceeds
the combined mass of the proton and electron. Therefore, asymmetric matter
with an admixture of electrons rather than pure neutron matter, is a more
likely composition of matter in neutron star interiors. The concentrations
of neutrons, protons and electrons can be determined from the condition of 
$\beta-$equilibrium 
$n\leftrightarrow p+e+{\bar{\nu}}$ 
and from charge neutrality, assuming that neutrinos are not degenerate. Here 
n, p, e, $\nu$ are have usual meaning as neutron, proton, electron and neutrino.
In momentum conservation condition 
$\nu_n=\nu_p+\nu_e,$ $n_p=n_e$, where 
$\nu_n=\mu_n-g_{\omega}V_0 + \frac{1}{2}g_{\rho}b_0$ and 
$\nu_p=\mu_p-g_{\omega}V_0 - \frac{1}{2} g_{\rho}b_0$ with 
$\mu_n={\sqrt{({k_{fn}^2}+{M^*{^2}_n})}}$ and 
$\mu_p={\sqrt{({k_{fp}^2}+{M^*{^2_p}
})}}$ are the chemical potential, and $k_{fn}$ and $k_{fp}$ are the Fermi
momentum for neutron and proton, respectively.
Imposing this conditions, in
the expressions of ${\cal E}$ and ${\cal P}$ (Eqns. 
\ref{eqn:eos1} - \ref{eqn:eos2}), we evaluate
 ${\cal E}$ and ${\cal P}$ as a function of density.
To calculate the star structure, we use
the Tolman-Oppenheimer-Volkoff (TOV) equations for the structure of a
relativistic spherical and
static star composed of a perfect fluid were derived from Einstein's
equations \cite{tov}, where the pressure and energy densities obtained from
equations (\ref{eqn:eos1}) and (\ref{eqn:eos2}) are the inputs.
The TOV equation is given by  \cite{tov}:

\begin{equation}\label{tov1}
\frac{d{\cal P}}{dr}=-\frac{G}{r}\frac{\left[{\cal E} + \cal P\right ]
\left[M+4\pi r^3 \cal P \right ]}{(r-2 GM)},
\end{equation}

\begin{equation}\label{tov2}
\frac{dM}{dr}= 4\pi r^2 \cal E,
\end{equation}
\begin{figure}
\caption{\label{fig9}The mass and radius of neutron star at different values of 
$g_\delta$. (a) $M/M_\odot$ with neutron star density (gm/{cm$^3$}), 
(b) $M/M_\odot$ with neutron star radius (km).}
\includegraphics[width=1.0\columnwidth]{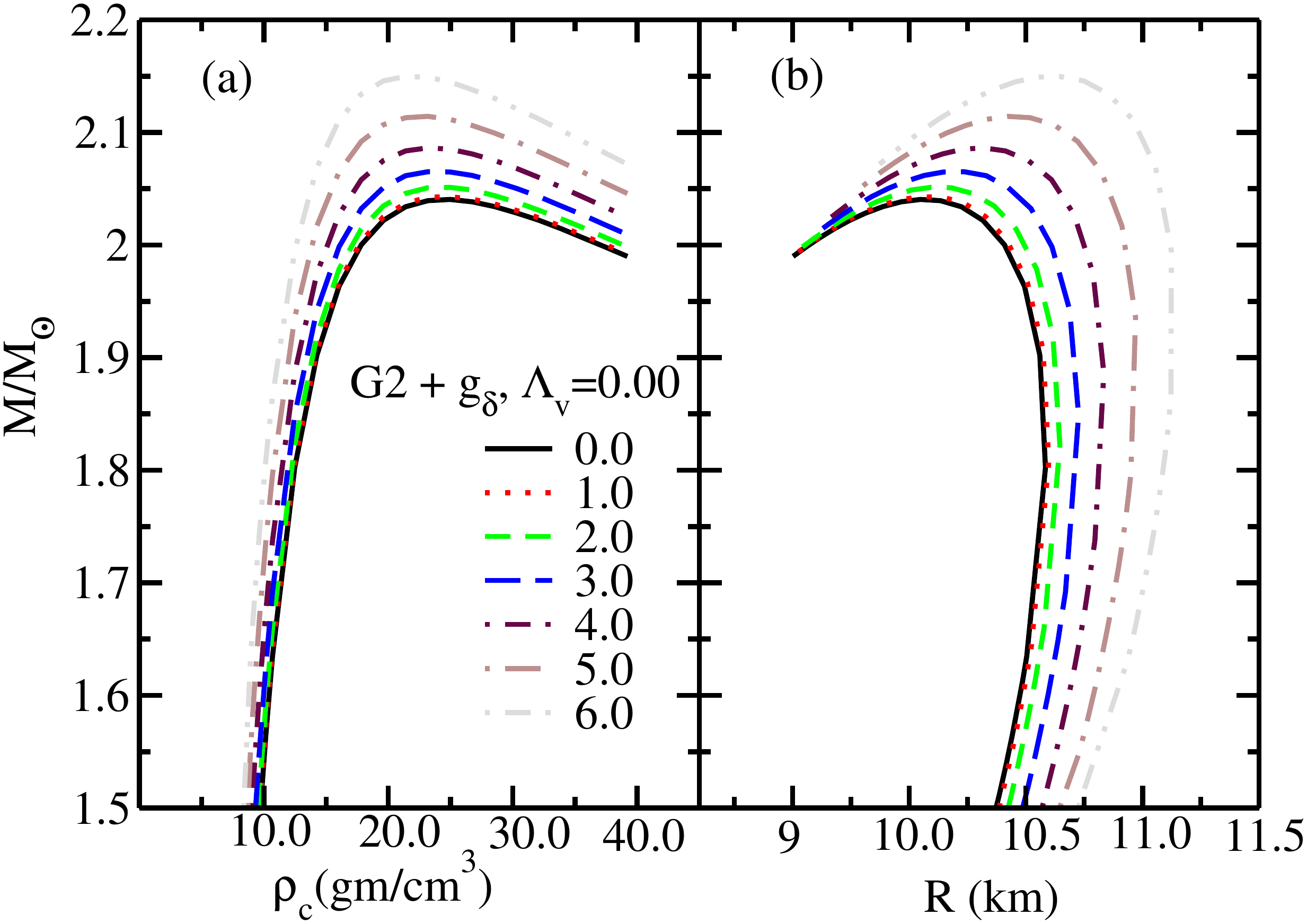}
\end{figure}
with $G$ as the gravitational constant and $M(r)$ as the enclosed
gravitational mass. We have used $c=1$.
Given the ${\cal P}$ and ${\cal E}$, these equations can be integrated
from the origin as an initial
value problem for a given choice of central energy density, $(\varepsilon_c)$.
The value of $r~(=R)$, where the pressure vanishes defines the
surface of the star.

The results of mass and radius with various $\delta-$meson coupling
strength $g_{\delta}$ is shown in Fig. \ref{fig9}. In left panel, the 
neutron star mass with density (gm/$cm^3$) is given, where we can see the 
effect of the 
newly introduced extra degree of freedom $\delta-$meson into the system. 
On the right side of the figure, [Fig. \ref{fig9}], $M/M_\odot$ is 
depicted with 
respect to radius (km), where M is the mass of the star and $M_\odot$ is the
solar mass. The $g_\delta$ coupling changes the star mass 
by $\sim$5.41\% and radius by $5.39\%$ with a variation of $g_{\delta}$ from
0 to 6.0. From this observation, we can say that 
$\delta-$meson is important not only for asymmetry system normal
density, but also substantially 
effective in high density system.
If we compare this results with the previous results \cite{shailesh13}, i.e., with the effects of cross 
coupling of $\omega-\rho$ on mass and radius of neutron star, the effects are
opposite to each other. That means, the star masses decreases with $\Lambda_v$, whereas
it is increases with $g_{\delta}$.
% We found that, 
%if we increase the coupling strength $\Lambda_v$, then mass and radius become 
%decrease and it increase with increasing the value of $g_\delta$.
Thus a finer tuning in mass and radius of neutron star is possible by a
suitable adjustment on $g_\delta$ value in the extended parametrization of
$G2+\Lambda_v + g_\delta$ to keep the star properties within the recent 
experimental observations \cite{demorest2010}.

\section{SUMMARY AND CONCLUSIONS}
In summary, we rigorously discussed the effects of cross coupling of 
$\omega-\rho-$mesons in finite nuclei on top of the pure G2 parameter set. 
The variation of binding energy, rms radii and energy levels of protons and
neutrons are analyzed with increasing values of $\Lambda_v$. The change in
neutron distribution radius $r_n$ with $\Lambda_v$ is found to be substantial
compared to the less effectiveness of binding energy and proton distribution radius
for the two representative nuclei $^{48}$Ca and $^{208}$Pb. Thus, to fix the
neutrons distribution radius depending on the outcome of PREX experimental 
\cite{jlab}
result, the inclusion of $\Lambda_v$ coupling strength is crucial. As it is 
discussed widely by various authors \cite{shailesh13}, the role of
$\omega-\rho-$mesons in the nuclear matter system is important on nuclear 
equation of states.

We emphasized strongly the importance of the effect of 
the extra degree of freedom, i.e., $\delta-$meson coupling into 
the standard RMF or E-RMF model, where, generally it is ignored. 
We have seen the effect of this coupling
strength of $\delta-$meson with nucleon in finite and neutron matter is substantial
and very different in nature, which may be extremely helpful to fix various
experimental constraints.  For example, with the help of $g_\delta$, it is 
possible to modify the 
binding energy, charge radius and flipping of the orbits in asymmetry finite nuclei 
systems.
The nuclear equation of state can be made stiffer  with the inclusion of
$\delta-$meson coupling. On the other hand, softening of symmetry energy is 
also possible with the help of this extra 
degree of freedom. 
In compact system, it is possible to fix the limiting values of 
$g_\delta$ and $\Lambda_v$ by testing the effect on available 
constraints on symmetry energy and its first derivative with respect 
to the matter density. 
This coupling may be extremely useful to fix the mass and radius
of neutron star keeping in view of the recent observation \cite{demorest2010}.

\newpage
%\section*{ACKNOWLEDGMENTS}

\end{document}